\documentclass[sigplan,screen]{acmart}
\usepackage{tikz}
\newcommand*\circled[1]{
    \tikz[baseline=(char.base)]{
        \node[shape=circle,fill,inner sep=0pt](char){
            \textcolor{white}{#1}
        };
    }
}

\DeclareMathOperator{\relu}{ReLU}

\usepackage{algorithm}
\usepackage[noend]{algpseudocode}

\usepackage{subfig}

\renewcommand\footnotetextcopyrightpermission[1]{} 
\def\BibTeX{{\rm B\kern-.05em{\sc i\kern-.025em b}\kern-.08emT\kern-.1667em\lower.7ex\hbox{E}\kern-.125emX}}
    
\copyrightyear{2019} 
\acmYear{2019} 
\setcopyright{acmlicensed}
\acmConference[ASPLOS '19]{2019 Architectural Support for Programming Languages and Operating Systems}{April 13--17, 2019}{Providence, RI, USA}
\acmBooktitle{2019 Architectural Support for Programming Languages and Operating Systems (ASPLOS '19), April 13--17, 2019, Providence, RI, USA}
\acmPrice{15.00}
\acmDOI{10.1145/3297858.3304048}
\acmISBN{978-1-4503-6240-5/19/04}

\begin{document}

%
\title[Field Programmable Synapse Array]{FPSA: A Full System Stack Solution for Reconfigurable ReRAM-based NN Accelerator Architecture}

%
\author{Yu Ji}
\email{jiy15@mails.tsinghua.edu.cn}
\affiliation{
    \institution{Tsinghua University}
    \city{Beijing}
    \country{China}
}

\author{Youyang Zhang}
\affiliation{
    \institution{Tsinghua University}
    \city{Beijing}
    \country{China}
}

\author{Xinfeng Xie}
\affiliation{
    \institution{University of California}
    \city{Santa Barbara}
    \country{USA}
}

\author{Shuangchen Li}
\affiliation{
    \institution{University of California}
    \city{Santa Barbara}
    \country{USA}
}

\author{Peiqi Wang}
\affiliation{
    \institution{Tsinghua University}
    \city{Beijing}
    \country{China}
}

\author{Xing Hu}
\affiliation{
    \institution{University of California}
    \city{Santa Barbara}
    \country{USA}
}

\author{Youhui Zhang}
\authornote{Corresponding author}
\email{zyh02@tsinghua.edu.cn}
\affiliation{
    \institution{Tsinghua University}
    \city{Beijing}
    \country{China}
}

\author{Yuan Xie}
\email{yuanxie@ece.ucsb.edu}
\affiliation{
    \institution{University of California}
    \city{Santa Barbara}
    \country{USA}
}
%
\renewcommand{\shortauthors}{Yu Ji, et al.}

%
\begin{abstract}
Neural Network (NN) accelerators with emerging ReRAM (resistive random access memory) technologies have been investigated as one of the promising solutions to address the \textit{memory wall} challenge, due to the unique capability of \textit{processing-in-memory} within ReRAM-crossbar-based processing elements (PEs).
However, the high efficiency and high density advantages of ReRAM have not been fully utilized due to the huge communication demands among PEs and the overhead of peripheral circuits.

In this paper, we propose a full system stack solution, composed of a reconfigurable architecture design, Field Programmable Synapse Array (FPSA) and its software system including neural synthesizer, temporal-to-spatial mapper, and placement \& routing.
We highly leverage the software system to make the hardware design compact and efficient.
To satisfy the high-performance communication demand, we optimize it with a reconfigurable routing architecture and the placement \& routing tool.
To improve the computational density, we greatly simplify the PE circuit with the spiking schema and then adopt neural synthesizer to enable the high density computation-resources to support different kinds of NN operations. 
In addition, we provide spiking memory blocks (SMBs) and configurable logic blocks (CLBs) in hardware and leverage the temporal-to-spatial mapper to utilize them to balance the storage and computation requirements of NN.

Owing to the end-to-end software system, we can efficiently deploy existing deep neural networks to FPSA. 
Evaluations show that, compared to one of state-of-the-art ReRAM-based NN accelerators, PRIME, the computational density of FPSA improves by $31\times$; for representative NNs, its inference performance can achieve up to $1000\times$ speedup.
\end{abstract}

\maketitle

\section{Introduction}\label{sec:introduction}

Neural Networks (NNs) have achieved state-of-the-art performance benefits in a wide range of AI applications ~\cite{krizhevsky2012alexnet,szegedy2015googlenet,simonyan2014vgg,he2016resnet,amodei2016deepspeech,sutskever2014seq2seq}, motivating the intensive studies on the design of NN accelerators to execute NN applications more efficiently.
ReRAM-based NN accelerator designs have been investigated as promising solutions due to the unique capability of performing efficient neural computing operations within ReRAM arrays~\cite{chi2016prime,hu2016dpe,shafiee2016isaac,song2017pipelayer}, which is called \textit{computing-in-memory} or \textit{processing-in-memory} (PIM) architecture enabled by the analog computing capability of ReRAM~\cite{hu2016dpe}.
Existing ReRAM-based NN accelerators~\cite{chi2016prime,shafiee2016isaac,song2017pipelayer} have shown a significant speedup over their digital counterparts~\cite{chen2014diannao,chen2014dadiannao,du2015shidiannao,chen2016eyeriss} because ReRAM can integrate computation and memory in the same physical place, which reduces the data movement between memory and computing elements.
ReRAM cells provide extremely high efficiency for dot-product computation, at high area density.
It takes approximately $10ps$\footnote{It is the resistive-capacitive delay of just the crossbar circuits} for a $100\times100$ crossbar~\cite{xia2016technological} to complete the vector-matrix multiplication
The size of an ReRAM cell is approximately $4F^2$~\cite{dong2014nvsim}, where $F$ is the feature size of the integrated circuit process.

Existing ReRAM-based NN accelerators usually use ReRAM-crossbar as the basic building block to calculate analog vector-matrix multiplication, and put a lot of efforts on hardware design to enable NN computation.
However, existing accelerators demonstrate far less efficiency and density than ReRAM's potential. The main bottleneck is communication.

\textbf{Communication Bottleneck.}
Without loss of generality, by the analysis (details are given in Section~\ref{sec:motivation}) of one of the state-of-the-art ReRAM-based NN accelerators, PRIME~\cite{chi2016prime}, we found that as the performance of processing elements (PEs) is increased significantly by ReRAM-crossbars, the communication between these PEs becomes a new system bottleneck.
Existing studies either use a memory bus~\cite{chi2016prime,song2017pipelayer} or Network-on-Chip (NoC)~\cite{shafiee2016isaac,liu2015reno} for communication.
The shared memory bus will inevitably become a bottleneck under the huge demand for data movement between PEs.
For NoC, the transmission latency is usually high and the bandwidth is still not enough for ReRAM-based PEs.


The analysis further shows that even if we solve the communication bottleneck, the overhead of peripheral circuits still makes the real performance of PE far from potential.

\textbf{Peripheral Circuit Overhead.}
Although ReRAM provides extremely high density, its peripheral circuits, such as analog-to-digital converters (ADCs) and digital-to-analog converters (DACs), occupy the majarity of a PE's area and processing latency, which seriously offsets the efficiency and density advantages.
Some recent studies~\cite{shafiee2016isaac,song2017pipelayer} try to reduce the overheads, but, fundamentally, the issue is not solved.
In addition, ReRAM crossbar is efficient when calculating vector-matrix multiplication. To support various and quickly evolving NNs, the peripheral circuits need to be more versatile in order to process a variety of operations, which worsens the problem.

To conquer these challenges, we propose an end-to-end full stack solution, which highly leverages software to use hardware resources efficiently, rather than complicating hardware. It is composed of a novel reconfigurable architecture for ReRAM-based NN accelerator, Field Programmable Synapse Array (FPSA), and the software system including neural synthesizer, spatial-temporal mapper, and placement \& routing.

For communication, we optimize the communication subsystem with a reconfigurable routing architecture, which provides massive
wiring resources for extremely high bandwidth and low latency and utilize them with the placement \& routing tool. Due to this optimization, we can achieve about two-orders-of-magnitude speedup in comparison of PRIME.

For peripheral circuits, we employ spiking schema to simplify the PE circuit while still maintaining the functionality of vector-matrix multiplication and Rectified Linear Unit (ReLU) activation  for artificial neural network (ANN).
We leverage the neural synthesizer to make the NN computation more compact and enable our high density homogeneous hardware to support different kinds of NN operations in order to fully utilize the advantage of ReRAM.
The latency and area of the entire PE is reduced by $94.90\%$ and $36.63\%$ respectively, which provides another order-of-magnitude speedup.

Last but not least, we introduce spiking memory blocks (SMBs) and configurable logic blocks (CLBs) in hardware as on-chip buffer and programmable logic. They are utilized by the spatial-to-temporal mapper to achieve optimized resource allocation and scheduling in order to balance the storage and computation requirements of NN, especially catering to the weight sharing property of convolutional neural networks (CNNs).
It can lead to super-linear performance increase with more hardware resources.

In our design, the performance is no longer bounded by the \textit{communication bottleneck}, and the \textit{peripheral circuit overhead} is significantly reduced. Experiments show that the performance is increased by $1000\times$ compared to PRIME~\cite{chi2016prime}, which is all due to the architectural and system improvements.

ReRAM-device variation is also considered: We propose a novel weight representation method, the \textit{add} method, to decrease device variation exposed to NN models. It can approach the full precision accuracy for large-scale NNs.

The contributions of this paper are summarized as follows.
\begin{itemize}
    \item We propose a full stack solution for ReRAM-based NN accelerator, including a reconfigurable architecture, FPSA, and the software hierarchy. The latter fully utilizes the various kinds of programmable resources provided by the former to deploy NN efficiently. Evaluations show that our approach can outperform a state-of-the-art ReRAM-based accelerator, PRIME, by up to $1000\times$ for NN inference.
    \item We have observed that communication is the bottleneck of existing ReRAM-based NN accelerator and then propose to optimize it with a reconfigurable routing architecture to break this bound.
    \item We make the PE design much more compact and efficient by leveraging the spiking schema. The latency is decreased by $19.6\times$ and the density is improved by $1.6\times$.
\end{itemize}

Finally, we believe that it is a new design philosophy for ReRAM-based NN accelerators.
Inspired by the spirit of the reduced instruction set computer (RISC) architecture of the conventional computer systems, our compact hardware design enables extremely high performance and can support rich NN functionalities with the software stack.


\section{Background and Related Work}\label{sec:background}

\subsection{ReRAM-Based NN Acceleration}\label{sec:reram_basis}

Neural Network applications are both memory-intensive and compute-intensive.
Thus, there are a lot of NN accelerators~\cite{chen2016eyeriss,chen2014diannao,chen2014dadiannao,du2015shidiannao,liu2015pudiannao,albericio2016cnvlutin,han2016eie,reagen2016minerva,kim2016neurocube,likamwa2016redeye,jouppi2017tpu} based on mature digital circuits to speedup NN computations.

To further increase the performance and eliminate other problems such as \textit{memory wall}, quite a few studies on ReRAM based NN accelerators and neuromorphic hardware~\cite{chi2016prime,shafiee2016isaac,song2017pipelayer,liu2015reno,jo2010nanoscale,pershin2010experimental,prezioso2015training,kim2012neural,indiveri2013integration,thomas2013memristor,adhikari2012memristor} have also been proposed.

\begin{figure}
    \centering
    \includegraphics[width=0.35\textwidth]{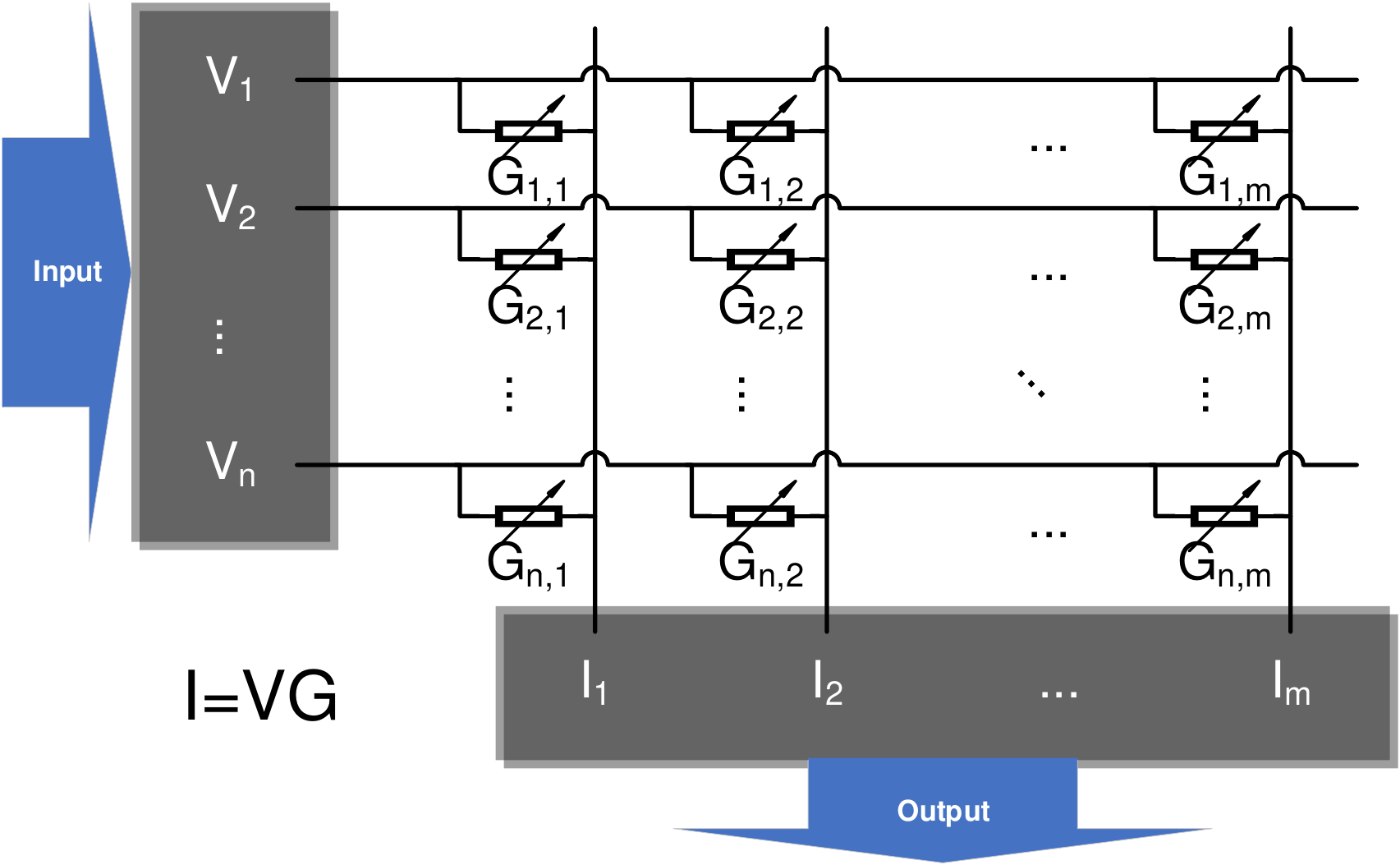}
    \caption{Vector-matrix multiplication with ReRAM crossbar}
    \label{fig:reram_crossbar}
\end{figure}

Resistive random access memory, known as ReRAM, is a type of emerging non-volatile memory, which stores the information using its resistance.
Prior work \cite{hu2016dpe} shows that the ReRAM-based crossbar is very efficient at computing analog vector-matrix multiplications in the locations where the matrices are stored.
As shown in Figure~\ref{fig:reram_crossbar}, there is a ReRAM cell in each intersection of the crossbar.
An input voltage vector $\{V_i\}$ is applied to the rows and is multiplied by the conductance matrix of ReRAM cells $\{G_{ji}\}$.
The resulting currents $\{I_j\}$ are summed across each column.
The output current vector can be calculated by $I = G V$.

Existing studies on ReRAM-based NN accelerators~\cite{chi2016prime,shafiee2016isaac,song2017pipelayer} treat the ReRAM-crossbar as a very low-precision vector-matrix multiplication engine, and use it as the building block, combined with peripheral circuits, to construct NN accelerators.
To support higher precision, these studies usually use the \textit{splicing} method, which employs multiple cells for different bits of the high precision number and shift-add the partial sum of different bits to get the final result.
For example, ISAAC conservatively uses 8 cells to represent one 16-bit cells; each cell represents 2 bits.
PRIME~\cite{chi2016prime} and PipeLayer~\cite{song2017pipelayer} are modified from the ReRAM-based memory chip.
Thus, their PEs are connected through the internal hierarchical memory bus.
ISAAC~\cite{shafiee2016isaac} is a dedicated accelerator, which employs NoC.

\subsection{Reconfigurable Architecture}\label{sec:reconfigurble_arch}
Reconfigurable architecture provides much higher efficiency than general-purpose processors while providing more flexibility than Application Specific Integrated Circuits (ASICs).
There are also some reconfigurable routing architectures designed for NN accelerators such as MAERI~\cite{kwon2018maeri}, but they target to the accelerators based on digital circuits. The capability is still far from the demands for ReRAM-based PEs.

FPGA is one of the most widely-used reconfigurable architectures, composed of many Configurable Logic Blocks (CLBs).
The main function modules of CLB are Look-Up Tables (LUTs) that can be configured to achieve any arbitrary logic function.
The routing architecture of an FPGA chip occupies up to $90\%$ of the total area~\cite{george2000fpga}, and provides most of the reconfigurability.
It consists of wires and programmable switches.
The programmable switches use Connection Boxes (CBs) to configure the connection from CLBs to the routing network, and use Switch Boxes (SBs) to configure the connections from different wire segments.
There have been many studies~\cite{xia2009memristor,cong2011mrfpga,wang2010fpga,owlia2014novel,vourkas2016digital,zha2018liquid} on using ReRAM to augment existing reconfigurable architectures.
For example, ReRAM cells are used to replace SBs and CBs in FPGA~\cite{cong2011mrfpga} and to implement arbitrary logic function~\cite{zha2018liquid}.

\section{Motivation}\label{sec:motivation}


\begin{figure}
    \centering
    \includegraphics[width=0.4\textwidth]{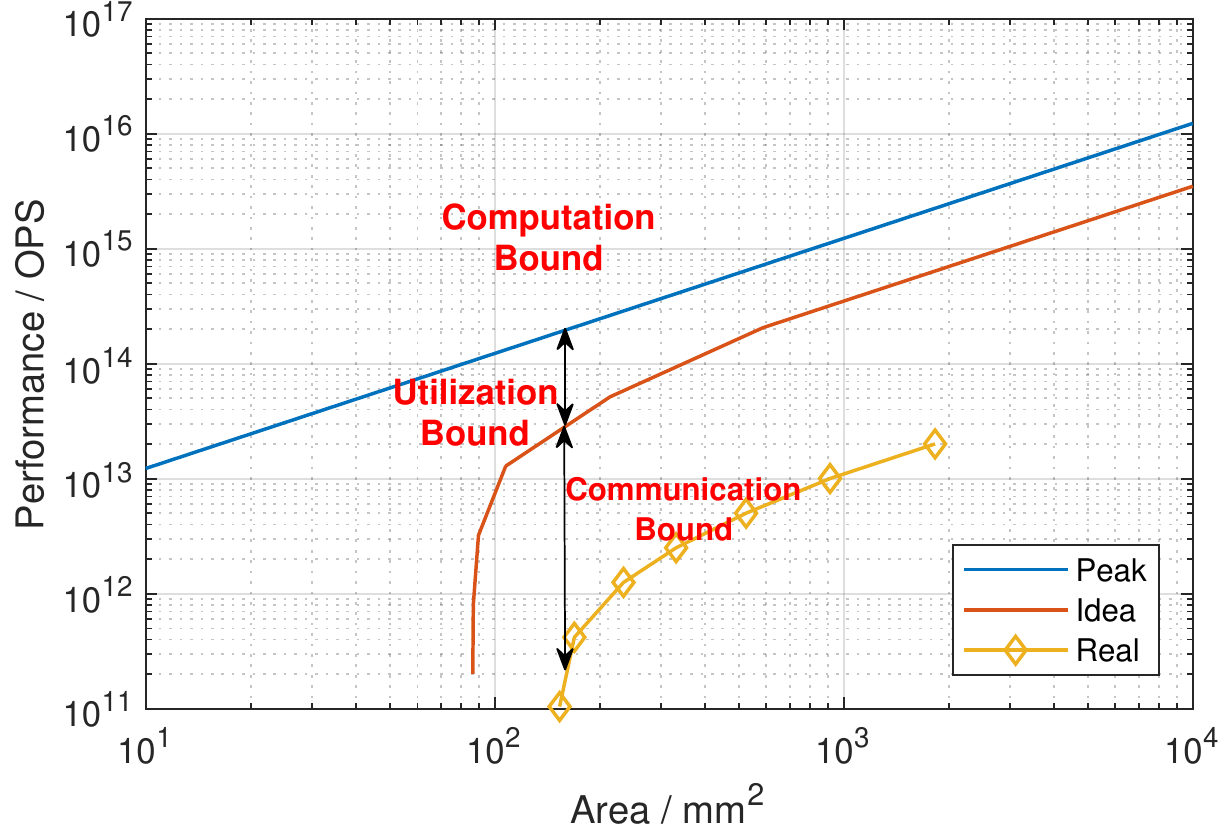}
    \caption{
        Performance vs. Area for the peak performance, the ideal case (with infinite bandwidth), and the real case for running VGG16~\cite{simonyan2014vgg} on PRIME~\cite{chi2016prime} ($45nm$ process). The performance of the real case is bounded by communication.
    }
    \label{fig:prime_performance}
\end{figure}

We analyze the scalability and performance of PRIME~\cite{chi2016prime}\footnote{Thanks to the authors of PRIME. We got all of its implementation code.}, which uses memory bus as the communication subsystem; we assume that its structure can scale-out linearly under 45nm process. A large scale CNN, VGG16~\cite{simonyan2014vgg} for ImageNet~\cite{deng2009imagenet}, is employed as the NN application.

Based on the hardware configurations and NN requirements, we can get three performance bounds (Figure~\ref{fig:prime_performance}) as follows. 



\textbf{Computation Bound.} 
It is the theoretical upper bound (which is defined as \textit{peak performance} in this paper), the product of the PE number and the performance of one PE, as the total computation capability provided.

\textbf{Utilization Bound.}
Usually, computation and communication capabilities are two important factors restricting performance improvement. But, even if the communication is ideal, the performance (called \textit{ideal performance}) still cannot reach the peak value, caused by the following two utilization issues:

\noindent$\bullet$ \textit{Temporal Utilization (Load Balance).} The first is the imbalance between storage and computation requirements of NN, especially for convolutional neural networks (CNNs).
For example, the first two convolutional layers of VGG16 only occupy $0.028\%$ of weight storage but consume $12.5\%$ of computation because the weights are reused by $224\times224$ different regions of the input feature map, while the fully connected layers take $89.3\%$ of storage but only consume $0.8\%$ of computation.

\noindent In contrast, ReRAM-crossbars integrate computation and storage in the same physical place; thus a PE can only provide computing power commensurate with its storage capacity.

\noindent To map a neural network onto the ReRAM-based NN accelerator, the prerequisite is that there should be enough PEs for all the weight parameters.
This mapping is quite unbalanced: about $0.028\%$ of PEs should process $12.5\%$ of computation and become the bottleneck, while the utilization of other PEs is low.
This issue can be solved when more PEs are available: We can duplicate these layers' weights onto more PEs to speedup them significantly.
For example, adding extra $0.028\%$ of PEs for the first two layers can double the performance.
That is why the first half of the \textit{ideal performance} curve shows a super-linear increase.
The curve will converge to linear scalability and approach the \textit{peak performance} when different layers are balanced.

\noindent$\bullet$\textit{Spatial Utilization (Crossbar Mapping).}
The fixed size of crossbars cannot match weight matrices of different scales perfectly, which also affects the PE utilization. 

\noindent Between the two, the first is the main issue.

\textbf{Communication Bound.}
In real cases with limited bandwidth, the utilization cannot be improved efficiently when more PEs are provided because the communication subsystem cannot fetch enough data in time for the PEs. 
This leads to a large gap with the ideal case.

Currently, PRIME has tried to balance the computation and communication requirements. However, due to its limited bus bandwidth, its real performance is far below the ideal value (two orders of magnitude lower than the latter). 

Based on these observations, it is reasonable to improve the performance of ReRAM-based accelerators with the following methods in order.
\begin{enumerate}
    \item \textbf{Improving Communication.} We should improve the communication subsystem to break the \textit{communication bound}.
    \item \textbf{Reducing Area.} We should reduce the area of a single PE to push the performance to the high-utilization region of the \textit{utilization bound} for a given chip area.
    \item \textbf{Reducing Latency.} We should reduce the latency of PEs to increase the peak performance (the \textit{upper bound)} further.
\end{enumerate}

Accordingly, we adopt the reconfigurable routing architecture first and then design simplified PE circuits to reduce area and latency, which are given in Section~\ref{sec:architecture}; the whole system software stack is proposed in Section~\ref{sec:system}. 
\begin{figure*}
    \centering
    \includegraphics[width=0.9\textwidth]{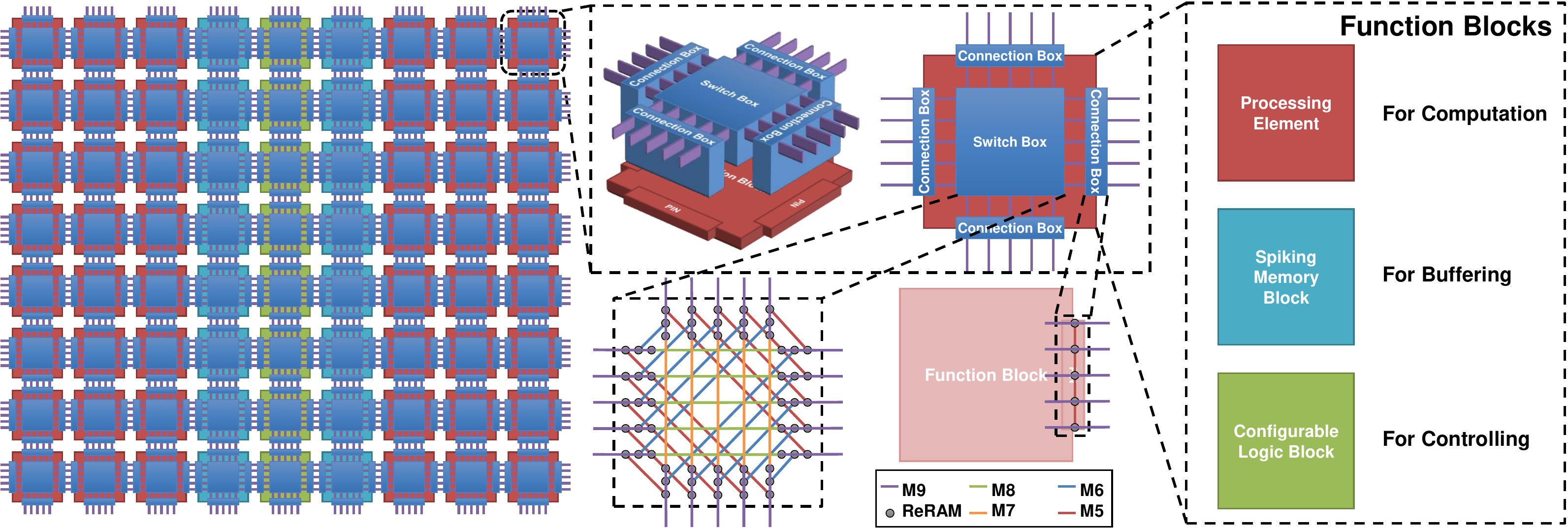}
    \caption{FPSA Architecture Overview. The function blocks are connected through the reconfigurable wiring network.}
    \label{fig:architecture}
\end{figure*}

\section{Architecture Design}\label{sec:architecture}


Figure~\ref{fig:architecture} shows the overview of FPSA architecture.
It contains three kinds of function blocks: ReRAM-based processing elements (PE) for computation, spiking memory block (SMB) for buffering, and configurable logic block (CLB) for controlling.
These blocks are connected through a reconfigurable routing architecture.
Functional blocks and the routing architecture are all programmable, which provide massive computation, buffering, controlling, and wiring resources for software to utilize.

To reduce the peripheral circuit overhead, we employ spiking schema to perform the vector-matrix multiplication.
It uses the spike count to represent a high-precision number rather than the amplitude of an analog signal.
The area and latency can be significantly reduced with this schema. 
In addition, the spiking memory block is customized to buffer spiking signals.

\subsection{Routing Architecture}\label{sec:routing_architecture}

PEs and other function blocks are connected by the routing architecture and working in parallel in a pipelined manner.
The pipeline clock cycle is bounded by the maximum latency of all pipeline stages, including the computation and communication latency.
As mentioned before, the computation time has been significantly reduced by ReRAM-crossbar, which makes the communication a system bottleneck.

Therefore, we adopt the reconfigurable routing architecture widely used in FPGA chips, instead of the memory bus or NoC in existing NN accelerators.
Compared to the memory bus and NoC that reuse physical channels for different traffic and provide flexible runtime data-path, the reconfigurable routing architecture assigns individual channels for each signal in the configuration phase and has a fixed runtime data-path (since the NN topology is fixed, the runtime flexibility is unnecessary).
Furthermore, compared to the bus and NoC where the worst communication latency is not guaranteed, the maximum latency of critical path can be evaluated in advance.

One of the most widely used FPGA routing architectures is the island-style architecture: configurable logic blocks (CLBs) are connected to the wiring network through connection boxes (CBs) and different wiring segments are connected through switch boxes (SBs).
Normally, the routing architecture consumes most of the FPGA chip area~\cite{george2000fpga}. In our design, the area consumption would be greater because of more fan-in/outs in the ReRAM-based PEs than those of CLBs in normal FPGA.

To reduce this overhead, we adopt the previous work, mrFPGA~\cite{cong2011mrfpga}, that employs ReRAM cells to construct CBs and SBs to reduce the area consumption.
Figure~\ref{fig:architecture} provides a detailed view of the routing architecture, in which SBs and CBs are placed over the function blocks.
Specifically, the connections in SBs and CBs are decided by the resistance of the ReRAM cells.
For example, an ReRAM cell with high resistance means that there is no connection between the two corresponding segments while low resistance is a pass.
Figure~\ref{fig:architecture} also provides the detailed wiring and layout inside CBs and SBs, which only use five metal layers from M5 to M9 without resource conflict.
Functional blocks are connected to the wiring network through the CBs at four sides.

\subsection{Processing Elements}\label{sec:processing_elements}

\begin{figure*}
    \centering
    \includegraphics[width=0.9\textwidth]{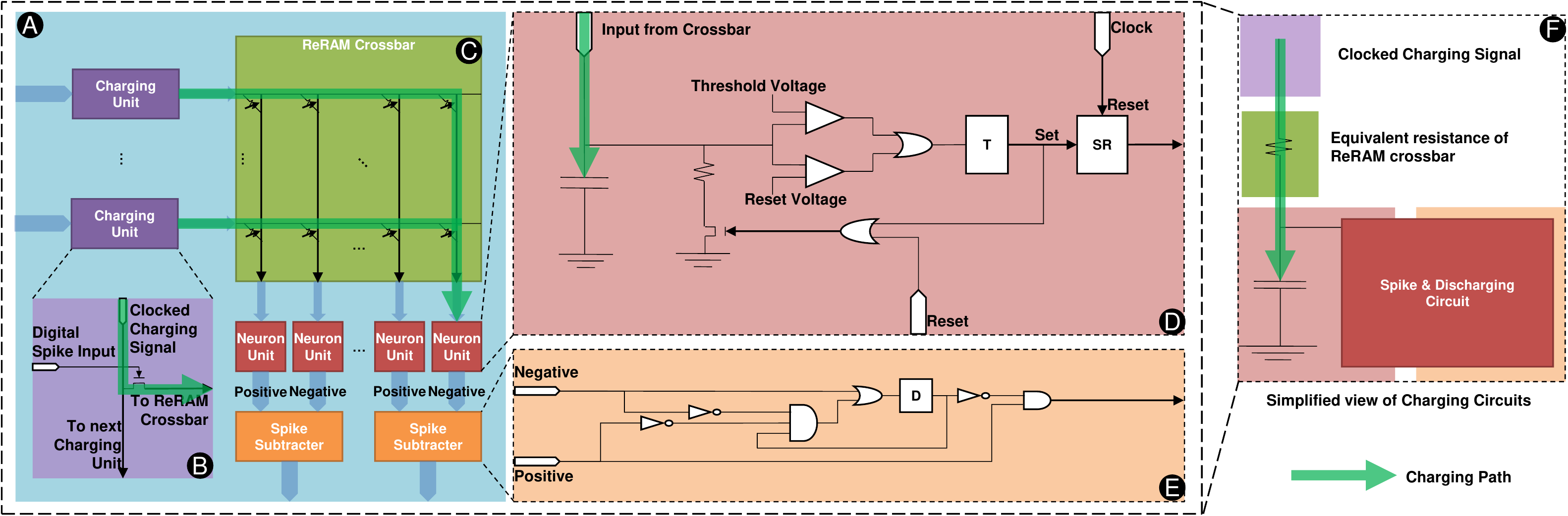}
    \caption{
        Overview of the Processing Element.
        The input is digital spike signals from routing architecture.
        The crossbar uses two columns for one output: one for the positive part and one for the negative.
        Neuron units integrate the output current from the corresponding crossbar-column and generate digital spikes.
        The spike subtracter computes the difference of the adjacent positive and negative columns.
        The green line represents the charging path of the capacitor of neuron unit.
        The simplified view of the charging circuits is on the right side of the figure.
    }
    \label{fig:processing_element}
\end{figure*}

We use spiking schema to simplify the peripheral circuits of PE.
The inputs of the PE are digital spike trains that use the spike count to represent a number between 0 and 1.
Although it requires $2^n$ spikes to represent a number of $n$ bits, processing spikes is much more efficient than processing high-precision analog signals comprehensively.

The essential of the PE is an ReRAM crossbar followed by spiking neuron circuits.
The input signal will be converted into a charging voltage and applied to each row of the crossbar.
Then the resulting current of each column will be injected into the corresponding neuron circuit, which accumulates the current and issues a spike when the threshold voltage is reached.

In order to handle negative weights with the positive conductance values, we use two physical adjacent columns to represent one logic column of the weight matrix, one for the positive part and one for negative.
The output spike train of the negative column will be subtracted from the positive one to get the final output.

Accordingly, the main components of a PE are charging units (one for each row), ReRAM-crossbar, neuron units (one for each column), and spike subtracters (one for every two columns).
The overview of a PE is shown in Figure~\ref{fig:processing_element}\circled{A}.

\textbf{Charging Unit.}
As shown in Figure~\ref{fig:processing_element}\circled{B}, since the input spike is a $1$-bit signal, the DAC can be simplified to a transistor.
When a spike signal arrives, the transistor will open and the charging voltage will be applied to this row.

\textbf{ReRAM Crossbar.}
Figure~\ref{fig:processing_element}\circled{C} is the ReRAM crossbar.
Each row connects to an input charging unit and each column connects to an output neuron unit.
ReRAM cells are in the intersections of the crossbar.

\textbf{Neuron Unit.}
It is an analog implementation of one widely used spiking neuron model, integrate-and-fire (IF) model.
As shown in Figure~\ref{fig:processing_element}\circled{D}, it has a capacitor to integrate the current from the corresponding column.
When its internal voltage reaches the threshold voltage, a spike signal will be stored in the S-R latch; the discharging unit will be turned on to discharge the capacitor until the voltage reaches the reset value.
The discharging unit can also be triggered by a reset signal because we use the spike count in a sampling window to represent a number. Thus, a reset signal will be sent to clear internal states before a new sampling window begins.

\textbf{Spike Subtracter.}
Figure~\ref{fig:processing_element}\circled{E} shows the circuit of the spike subtracter.
It has two input spike trains from the corresponding two neuron units.
The output is also a spike train, whose spike count is the different of the two inputs.
The working mechanism is that the spikes from the negative neuron unit will block the next spike coming from the positive neuron.

Although we use spiking schema in our circuit design, the computation achieved by the circuit is just a vector-matrix multiplication followed by the ReLU activation function; the precision depends on the size of the sampling window. 
The proof is as follows.
The equivalent charging circuit is shown in Figure~\ref{fig:processing_element}\circled{F}.
We denote the charging voltage from the voltage source as $V_{dd}$, the capacitance of neuron unit as $C$, and the charging time of each clock cycle as $\tau$.
For the $j$-th output neuron unit, the equivalent resistance of the ReRAM-crossbar is denoted as $R_j(t)$ at time $t$.
We suppose that from the reset voltage $V_{re}$, the neuron unit's capacitor reaches the threshold $V_{th}$ in the $T$-th cycle.
In accordance with the model of charging a capacitor in an RC circuit, Equation~\ref{eq:U_T} gives the capacitor's voltage $U_T$ at the cycle $T$.

\begin{equation}\label{eq:U_T}
    V_{dd} - U_T = (V_{dd} - U_{T-1})e^{-\frac{\tau}{R_j(T)C}} = (V_{dd} - V_{re}) e^{-\frac{\tau}{C}\sum_{t=1}^T \frac{1}{R_j(t)}}
\end{equation}

When $U_T$ reaches the threshold $V_{th}$ at the $T$-th cycle, we can derive Equation~\ref{eq:Rj_eta}.

\begin{equation}\label{eq:Rj_eta}
    \sum_{t=1}^T\frac{1}{R_j(t)} = \frac{C}{\tau}\ln{\frac{V_{dd} - V_{re}}{V_{dd} - V_{th}}}
\end{equation}

For convenience, we denote the right-hand side of Equation~\ref{eq:Rj_eta} as $\eta$ because it is a constant.
On the left-hand side, the equivalent resistance only counts the rows with spike inputs.
Therefore, we can derive Equation~\ref{eq:si_eta} as follows, where $s_i(t)$ is the spike signal for the $i$-th row at time $t$ and $g_{ji}$ is the conductance of the cell at the intersection of the $i$-th row and the $j$-th column.



\begin{equation}\label{eq:si_eta}
    \sum_{t=1}^T\frac{1}{R_j(t)} = \sum_{t=1}^T\sum_i s_i(t)g_{ji} = \sum_i g_{ji} \sum_{t=1}^T s_i(t) = \eta
\end{equation}

Suppose the size of the sampling window is $\Gamma$ cycles.
During this period, the spike counts of the $i$-th input row and the $j$-th output column are $X_i$ and $Y_j$ respectively.
Thus, the voltage of the capacitor has reached the threshold for $Y_j$ times and then we have Equation~\ref{eq:Yj_si}.

\begin{equation}\label{eq:Yj_si}
    \sum_i g_{ji} \sum_{t=1}^{\Gamma} s_i(t) = Y_j\eta
\end{equation}

By definition, $X_i$ is the sum of $s_i(t)$ of the sampling window $\Gamma$. Thus, the relationship between the input and output spike count is shown in Equation~\ref{eq:Yj_Xi}.

\begin{equation}\label{eq:Yj_Xi}
    Y_j = \sum_i \frac{g_{ji}}{\eta}X_i
\end{equation}

Further, we connect two columns to one spike subtracter to support negative weight values.
Suppose the corresponding spike counts and conductance values for positive and negative columns are $Y_j^+$, $Y_j^-$ and $g_{ji}^+$, $g_{ji}^-$, respectively.
The subtracter blocks $Y_j^-$ spikes from the $Y_j^+$ if $Y_j^+ > Y_j^-$, or the output spike count is $0$.
Thus the final spike count from the $j$-th output port is shown in Equation~\ref{eq:Yj_Xi_final}.

\begin{equation}\label{eq:Yj_Xi_final}
    Y_j = \max(Y_j^+ - Y_j^-, 0) = \relu(\sum_i\frac{g_{ji}^+ - g_{ji}^-}{\eta}X_i)
\end{equation}

In conclusion, the difference from existing ReRAM-based accelerators that employ spiking schema (e.g. PipeLayer~\cite{song2017pipelayer}) is that we directly charge the capacitor and transit spike trains between PEs. Thus, the overhead of current mirrors and encoder/decoder for spike trains can be removed.
Equation~\ref{eq:Yj_Xi_final} shows that with this simplification we can still complete the vector-matrix multiplication followed by ReLU.
In addition, owing to the area reduction, we do not need to reuse peripheral circuits for different rows and columns. They can process input and output of an ReRAM-crossbar in parallel. 
In contrast, existing ReRAM-based accelerators usually share ADCs and/or DACs to reduce the area overhead, which also leads to a corresponding increase in processing delay. (e.g. in ISAAC~\cite{shafiee2016isaac}, 128 crossbar-columns share one ADC).
Our approach achieves a good balance in terms of function, area cost and time delay. Quantitative evaluation will be given in Section~\ref{sec:evaluation}.

\subsection{Spiking Memory Block}
As shown in Figure~\ref{fig:architecture}, in addition to the computation resources provided by PEs, we also have spiking memory blocks (SMBs) to provide on-chip buffer for the intermediate data.

Since the size of on-chip buffers has a significant impact on chip area, we only store the spike counts instead of the spike trains to fully use the buffers.
The counters and spike generators are embedded inside the SMB to do the encoding and decoding between spike counts and spike trains; thus SMB can directly send and receive spike trains but only store the spike counts.
The internal memory is indexed by bits so that it can fit any sampling window size (e.g., when the sampling window is $2^n$, it can store the spike counts in the manner of $n$-bit by $n$-bit.

Although we heavily adopt ReRAM in our PE design and routing architecture, we still use SRAM for the SMB.
ReRAMs are not suitable for buffers because they have low endurance (they can support about $10^{12}$ writes).

\subsection{Configurable Logic Block}
Further, we provide configurable logic blocks (CLBs) to provide logic resources for controlling as shown in Figure~\ref{fig:architecture}.
The control signals for PEs and SMBs are generated by the CLBs.

We also use SRAMs to implement the LUTs in CLBs.
Although ReRAM provides higher density than SRAM, it requires current sense amplifiers to read data, which consume a lot of area.
Thus, its area efficiency is very poor when the capacity is small:
A conventional $6$-input LUT can be implemented with a $64$-bit memory.
According to NVSim~\cite{dong2014nvsim}, the area of a $64$-bit SRAM cell is $35.129 \mu m^2$ under $45 nm$ process while the area of an ReRAM cell is $172.229 \mu m^2$.
Thus, CLBs contain multiple SRAM-based LUTs, flip-flops, and multiplexers to perform any logic function.
\section{System Design}\label{sec:system}

\begin{figure*}
    \centering
    \includegraphics[width=0.9\textwidth]{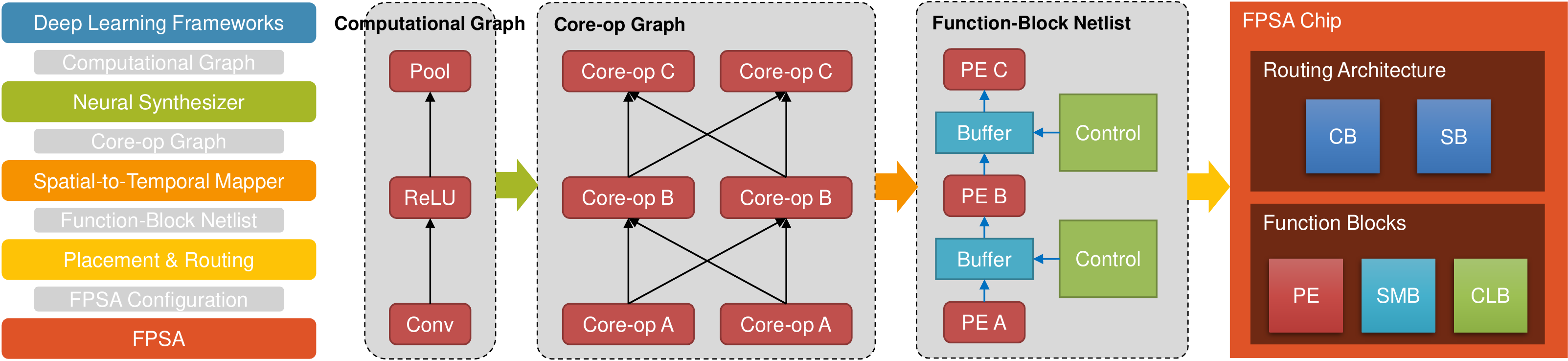}
    \caption{System stack of FPSA.}
    \label{fig:system}
\end{figure*}

We highly leverage the software system to enable flexible functionality and high efficiency of FPSA architecture.
Utill now, the hardware has provided massive computation, buffering, and controlling resources in the form of the three kinds of function blocks, as well as the massive wiring resources and configurable connections through the routing architecture. 
How to make full use of these hardware resources to fit the diversity of NN requirements is a complex problem, especially as we try to maintain the advantages of ReRAM (i.e. the high computational density of vector-matrix multiplication).

From a formal perspective, most deep learning frameworks~\cite{paszke2017pytorch,abadi2016tensorflow,chen2015mxnet} use computational graph (CG) as the programming model to represent NNs. Thus, the problem is how to efficiently map the software-level CG to the above reconfigurable resource pool. 

We divide the problem into three independent sub-problems and design the software stack to solve them respectively, as shown in Figure~\ref{fig:system}.
First, the neural synthesizer transforms the NN CG to make up the gap between the NN requirement and hardware functionality.
Second, the spatial-to-temporal mapper gives the optimized allocation of PE-resources and the scheduling strategy for the above-mentioned output CG, including the corresponding control logic; all of them are collectively referred to as the function-block netlist. 
Finally, we place the netlist onto the FPSA chip and generate the routing.

\subsection{Neural Synthesizer}
Here the essential is to maintain the user-friendly programming interface and synthesize NN model into a hardware-friendly, compact representation for efficient execution.

\textbf{Flexible NN Programming.}
Computational graph (CG) is widely used programming model in most deep learning frameworks.
It is a graph that consists of many tensor operations and describes the data dependencies of the operations.
There are hundreds of flexible and complex operations in most deep learning frameworks.


\textbf{Efficient ReRAM Execution.}
The support of hundreds of operations in hardware is impractical. On the other side, our ReRAM-based PE can complete vector-matrix multiplication with ReLU function very efficiently (in Section~\ref{sec:processing_elements}).
Therefore, the neural synthesizer is expected to synthesize the software CG into an equivalent CG only including operations that the hardware can support efficiently. 

We adopt the existing NN compiler framework from Y. Ji et al~\cite{ji2016neutrams,ji2018compiler} to do the synthesis. They propose to transform a trained, software NN into an equivalent network that meets hardware constraints; one case study is to transform such a CG into a core-op graph (core-op is defined as an operation composed of a low-precision vector-matrix multiplication and ReLU). Namely, it can implement different kinds of operations with the core-op, and then fine-tune the model to retain the accuracy.
The basic idea is to construct dedicated structures with core-ops to implement other operations or approximate them with multilayer perceptrons (MLPs).
Further, large fully-connected layers or convolutional layers will be split into multiple small core-ops.

\subsection{Spatial-to-Temporal Mapper}\label{sec:mapper}
The output core-op graph only contains purely computational tasks. If we map CG nodes onto PEs directly, it will require extremely huge amount of PEs, which is impractical.
For example, although a convolutional layer reuses its kernel weights for different regions of input feature map, its core-op graph contains individual core-ops for each region.
Thus, we have to temporally map the core-op graph onto hardware with the on-chip buffering and controlling resources. Still taking the convolutional layer as an example, we can map all core-ops with shared weights onto one or more PEs and reuse the weights in a time-division-multiplexing manner.
Accordingly, the mapper will generate an optimized netlist of function blocks for the core-op graph: PEs complete all the computation tasks, buffers hold the intermediate data, and control logic will be generated to schedule the execution.
Further, the buffers separate the entire circuits into multiple pipeline stages and different pipeline stages process different samples in parallel.
The mapping involves the following two sub-steps.

\textbf{Resource Allocation.}
As discussed in Section~\ref{sec:motivation}, different layers reuse the weights for different times.
We should assign more PEs to those layers that reuse weights more times.
To do so, we have all the core-ops with the same weights into one group. 
The number of core-ops in one group is denoted as \textit{reuse degree}.
The iterations required to complete the computation of a group depends on the number of PEs assigned to that group.
We first allocate one PE for each group to satisfy the minimum storage requirement.
To balance the pipeline stages, we will assign extra PEs to those groups that require more iterations to complete if more PEs are available.
The number of duplications (PEs) assigned to one group is referred as \textit{duplication degree} of that group.
We use the \textit{duplication degree} of the group with the maximum \textit{reuse degree} as the \textit{duplication degree} of the entire model.
With $n\times$ duplication degree, the temporal utilization bound is usually increased by $n\times$.

\textbf{Scheduling.}
After the core-ops are assigned to PEs, we also need to schedule the execution order, insert buffers between PEs, and generate the control signal to get the netlist.
We denote the core-op graph as $G=(V,E)$ where $V$ is the node set and $E$ is the edge set.
$A_v$ denotes the PE assigned to the core-op $v\in V$.
$s_v$ and $e_v$ represent the start cycle and end cycle for executing the core-op $v$ respectively.
The following contraints should be satisfied.

\noindent\textbf{$\bullet$ Resource Conflict (RC).} Two core-ops cannot be executed synchronously if they are assigned to the same PE, which is shown in Formula~\ref{eq:resource_conflict}.
\begin{equation}\label{eq:resource_conflict}
    e_v < s_u \mbox{ or } e_u < s_v \quad \mbox{if $A_v = A_u$}
\end{equation}

\noindent\textbf{$\bullet$ No-Buffer Dependency (NBD).}
If there is data dependency between node $u$ and $v$, and if these two nodes are placed into directly connected PEs without buffers, the execution time of $v$ needs to cover the one of $u$ to receive the spike train generated by $u$, as shown in Formula~\ref{eq:nobuffer_dependency}.
\begin{equation}\label{eq:nobuffer_dependency}
    s_v \le s_u + 1 \mbox{ and } e_v \ge e_u + 1 \quad \mbox{if $(u, v) \in E$}
\end{equation}

\noindent\textbf{$\bullet$ Buffered Dependency (BD).}
\textit{Resource conflict} and \textit{no-buffer dependency} may conflict; thus we add buffers between the two PEs to solve conflict.
The buffers will store the firing rate of $u$ and generate spikes for $v$ when $A_v$ is ready. This constraints is given by Formula~\ref{eq:buffered_dependency}.
\begin{equation}\label{eq:buffered_dependency}
    s_v > e_u \quad \mbox{if $(u,v)\in E$}
\end{equation}

\noindent\textbf{$\bullet$ Buffer Conflict (BC).}
If two nodes $u$ and $v$ receive spike trains from the same port of one buffer, the buffer should provide spike train of sampling window $\Gamma$ one-by-one.
The timing should satisfy Formula~\ref{eq:buffer_conflict}.
\begin{equation}\label{eq:buffer_conflict}
    e_v > e_u + \Gamma \mbox{ or } e_u > e_v + \Gamma
\end{equation}

\noindent\textbf{$\bullet$ Sampling Window (SW).}
Finally, the execution time of each core-op cannot be less than $\Gamma$ as Formula~\ref{eq:sampling_window}.
\begin{equation}\label{eq:sampling_window}
    s_v + \Gamma \le e_v
\end{equation}

We can optimize all the $s_v$ and $e_v$ for a certain objective under these constraints.
Here, we show a greedy algorithm in Algorithm~\ref{alg:schedule} to minimize the buffer used and the latency.
\begin{algorithm}
    \caption{Scheduling algorithm}\label{alg:schedule}
    \begin{algorithmic}
        \Require $G=(V,E)$, $A_v$
        \State $s_v, e_v$ is the start/end time of $v \in V$
        \For {$v \in V$ in topological ordering}
            \State Let $v$ satisfy \textit{NBD} and \textit{SW}
            \State Increase $s_v$, $e_v$ to satisfy \textit{RC}
            \If {$v$ does not satisfy \textit{NBD} with $u$}
                \State Mark $(u, v)$ requires buffer
                \State Increase $s_v$, $e_v$ to satisfy \textit{RC} and \textit{BD}
            \EndIf
            \For {$u$ where $(u,v)\in E$}
                \If {any $(u, p)$ requires buffer}
                    \State Insert buffer after $u$
                    \If {the buffer requires extending fan-in/out}
                        \For {$w$ where $(u, w)\in E$ requires buffer}
                            \State Increase $s_w$, $e_w$ to satisfy $BC$
                            \For {$q\in V$ between $w$ and $v$}
                                \State Increase $s_q$,$e_q$to satisfy all
                            \EndFor
                        \EndFor
                    \EndIf
                \EndIf
                \State Increase $s_u$, $e_u$ to satisfy all
                \For {$p\in V$ before $u$ in reverse ordering}
                    \State Increase $s_p$, $e_p$ to satisfy all
                \EndFor
            \EndFor
        \EndFor
    \end{algorithmic}
\end{algorithm}

The basic idea is to traverse the graph in topological ordering and try to connect PEs without buffer.
If there is any conflict, a buffer from SMB should be inserted to separate them into different pipeline stages.
Then we will check all the previous nodes and adjust them to ensure all constraints are satisfied.

When all $s_v$ and $e_v$ have been determined, the controlling signals can be generated accordingly with the CLBs.

\subsection{Placement \& Routing}
The last step is to place all function blocks of the netlist onto physical units.
Then the CBs and SBs in the routing architecture can be configured to connect the function blocks according to the topology of the netlist.
The placement \& routing problem is the same as the one for FPGA.
We adopt the mature solution used in FPGA development tool-chain, which usually uses simulated annealing (SA) algorithm for the placement, and uses dijkstra's shortest path algorithm for the routing to minimize the latency of critical path.

\section{Evaluation}\label{sec:evaluation}

\begin{table}
    \centering
    \caption{Parameters of function blocks under $45nm$ process}
    \label{tab:fpsa_parameters}
    \begin{tabular}{c r r r}
        \hline\hline
        & \textbf{Energy}   & \textbf{Area} & \textbf{Latency}\\
        & $pJ$              & ${\mu m}^2$   & $ns$\\
        \hline
        \textbf{PE ($256\times256$)}    & 29.094 & 22051.414 & 2.443\\
        \hline
        Charging Unit                   &  0.001 &     2.246 & 0.070\\
        $\times256$                     &  0.229 &   600.704 &      \\
        ReRAM ($256\times512$)          &  0.131 &  1061.683 & 0.000\\
        $\times8$                       &  1.049 &  8493.466 &      \\
        Neuron Unit                     &  0.039 &    19.247 & 1.463\\
        $\times512$                     & 19.861 &  9854.342 &      \\
        Subtractor                      &  0.031 &    12.121 & 0.910\\
        $\times256$                     &  8.945 &  3102.902 &      \\
        \hline
        \textbf{CLB ($128\times$ LUT)}  &  3.106 &  5998.272 & 0.229\\
        \hline
        \textbf{SMB ($16$Kb)}           &  1.150 &  5421.900 & 0.578\\
        \hline
    \end{tabular}
\end{table}

\begin{figure}
    \centering
    \includegraphics[width=0.45\textwidth]{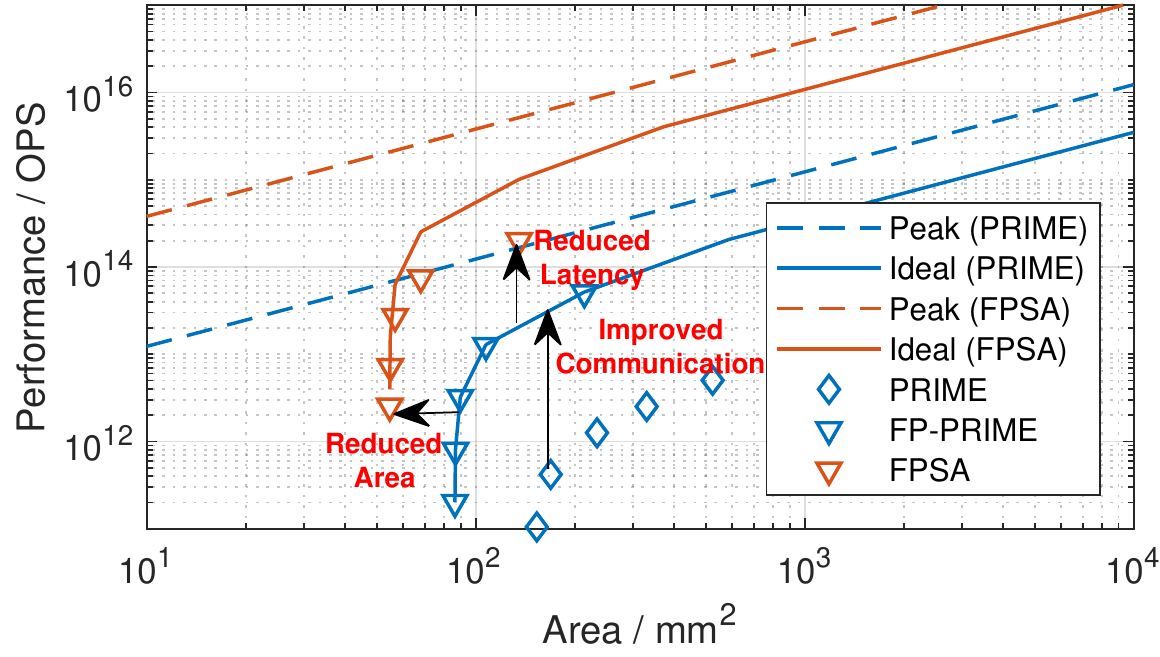}
    \caption{Comparison between PRIME, FP-PRIME (FPSA with PRIME's PE), and FPSA for VGG16.}
    \label{fig:overall_comparison}
\end{figure}

\begin{figure}
    \centering
    \includegraphics[width=0.45\textwidth]{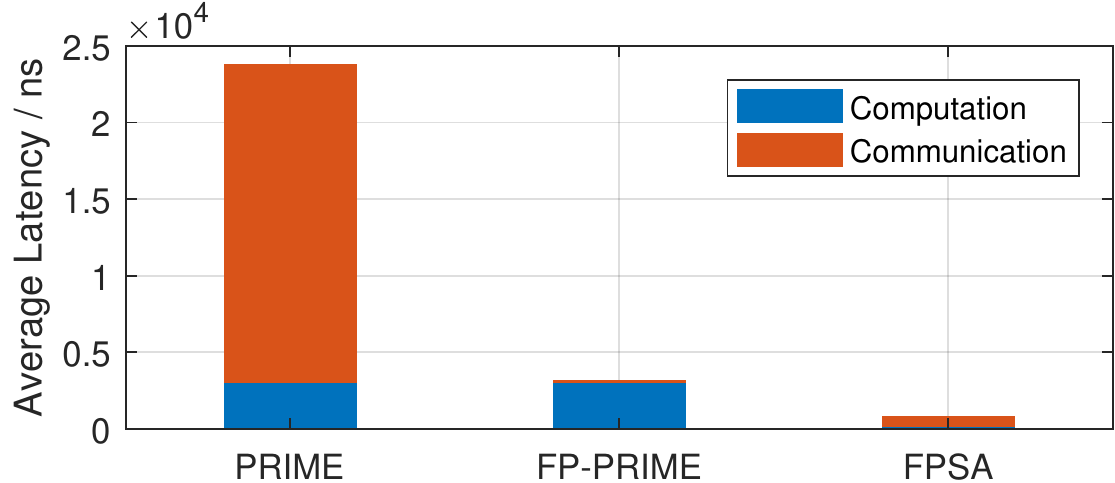}
    \caption{The breakdown of processing latency of one PE of PRIME, FP-PRIME, and FPSA (for VGG16).}
    \label{fig:latency_breakdown}
\end{figure}

\begin{table}
	\centering
	\caption{The comparison to PRIME for performing a vector-matrix multiplication of 8bit-weight, 6bit-I/O, and $256\times256$-scale.}
    \label{tab:ops_comprison}
    \begin{tabular}{c c c c c}
    	\hline\hline
        & \textbf{Area} & \textbf{Latency} & \textbf{Computational}\\
        & & & \textbf{Density}\\
        & ($\mu m^2$) & ($ns$) & ($OPS/mm^2$)\\
        \hline
        \textbf{PRIME}          & 34802.204 & 3064.7    & 1.229T\\
        \textbf{FPSA}           & 22051.414 & 156.4     & 38.004T\\
        \hline
        \textbf{Improvement}    & -36.63\% & -94.90\%  & $30.92\times$\\
        \hline
    \end{tabular}
\end{table}

\begin{figure*}
    \centering
    \subfloat{\includegraphics[width=0.33\textwidth]{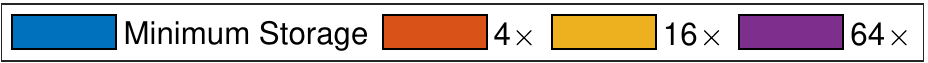}}
    
    \setcounter{subfigure}{0}
    
    \hspace*{\fill}
    \subfloat[Performance]{\label{fig:fpsa_performance}\includegraphics[width=0.3\textwidth]{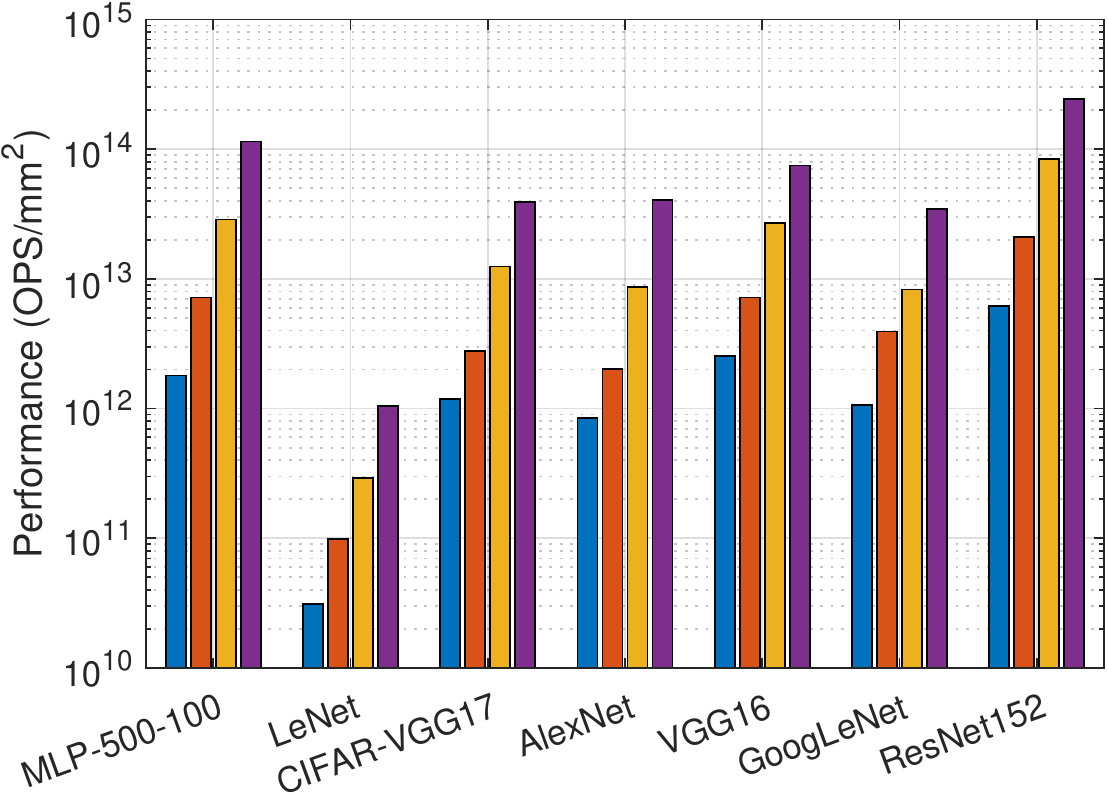}}
    \hspace*{\fill}
    \subfloat[Area]{\label{fig:fpsa_area}\includegraphics[width=0.3\textwidth]{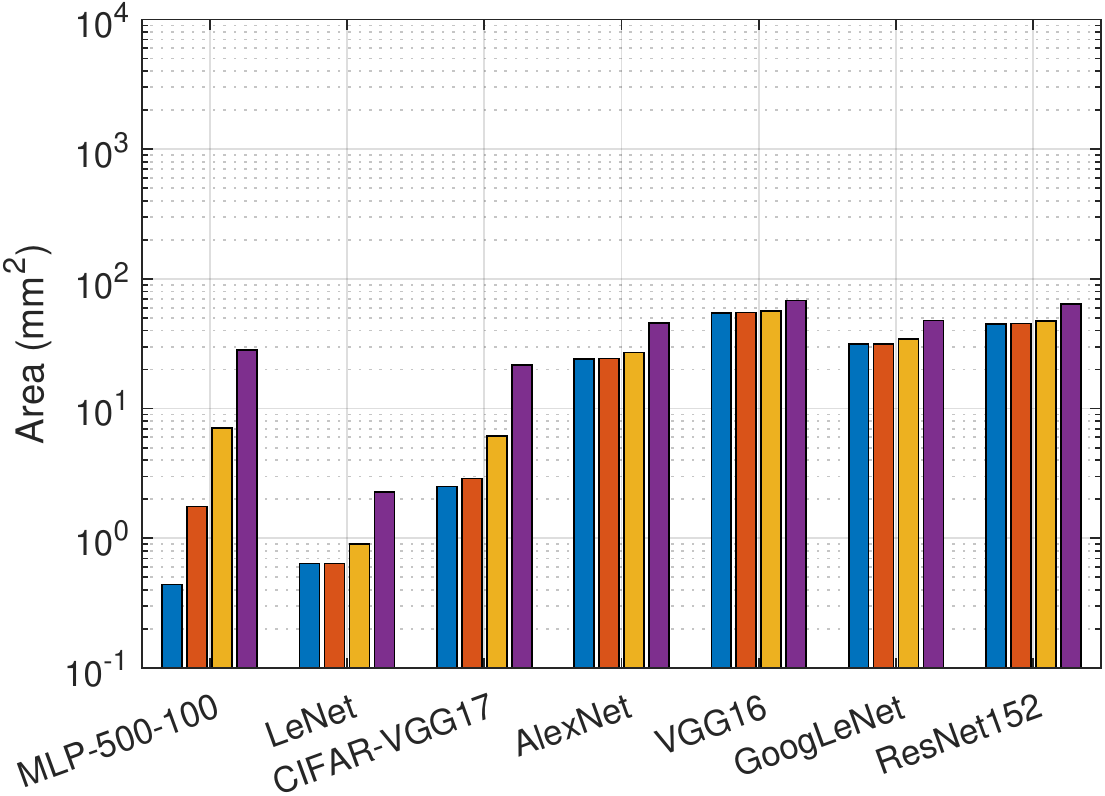}}
    \hspace*{\fill}
    \subfloat[Computational Density]{\label{fig:fpsa_utilization}\includegraphics[width=0.3\textwidth]{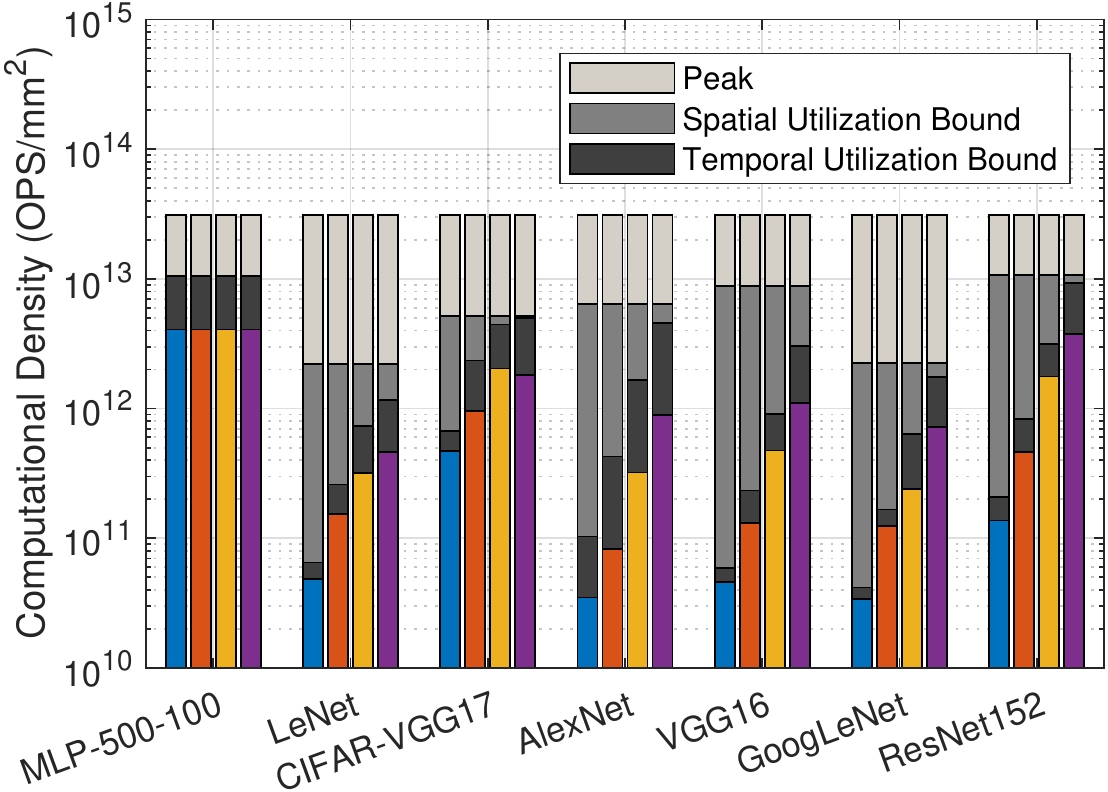}}
    \hspace*{\fill}
    
    \caption{Scalabilty of FPSA. We show the area, computational density, and performance for all the benchmark models under different \textit{duplication degrees}.
    (a) The performance increases significantly with the increase of \textit{duplication degree}. (b) The area consumption does not increase much as performance. (c) The rest performance-increase comes from the better utilization since the \textit{Temporal Utilization Bound} increases as more resources are available.}
    \label{fig:scalability}
\end{figure*}

\begin{table*}
    \centering
    \begin{tabular}{c r r r r r r r}
        \hline\hline
        \textbf{Model} & MLP-500-100 & LeNet & VGG17 & AlexNet & VGG16 & GoogleNet & ResNet152\\
        \hline
        \textbf{Dataset} & MNIST & MNIST & CIFAR-10 & ImageNet & ImageNet & ImageNet & ImageNet\\
        \hline
        \textbf{\# of weights} & 443.0K & 430.5K & 1.1M & 60.6M & 138.3M & 7.0M & 57.7M\\
        \textbf{\# of ops} & 886.0K & 4.6M & 333.4M & 1.4G & 30.9G & 3.2G & 22.6G\\
        \hline
        \textbf{Throughput ($sample/s$)} & 129.7M & 229.4K & 117.4K & 28.2K & 2.4K & 10.9K & 10.8K\\
        \textbf{Latency ($\mu s$)} & 0.51 & 0.97 & 46.3 & 100.49 & 671.8 & 514.18 & 1106.4\\
        \textbf{Area ($mm^2$, $45nm$ process)} & 28.23 & 2.27 & 21.68 & 45.89 & 68.09 & 47.74 & 64.32\\
    \hline
    \end{tabular}
    \caption{The overall performance of FPSA for different NN models}
    \label{tab:fpsa_overall}
    \vspace{-10pt}
\end{table*}

We evaluate the FPSA architecture and its system stack with a set of typical NN applications. Specifically, we evaluate the contributions of the routing architecture and simplified PEs to the whole system improvement separately.
Further, the scalability is evaluated when more resources are provided.

\subsection{Experiment Configurations and Methodology}

\textbf{Benchmark.}
We evaluate our proposal on NN models of different scales, including MLP-500-100 for MNIST dataset~\cite{lecun1998mnist} (an MLP with two hidden layers composed of 500 and 100 neurons), LeNet~\cite{lecun2015lenet} for MNIST dataset, VGG17 for CIFAR-10 dataset~\cite{krizhevsky2009cifar}, AlexNet~\cite{krizhevsky2012alexnet}, GoogLeNet~\cite{szegedy2015googlenet}, VGG16~\cite{simonyan2014vgg}, and ResNet152~\cite{he2016resnet}.
The last four are for the ImageNet dataset~\cite{deng2009imagenet}.

\textbf{Baseline.}
We compare FPSA to state-of-the-art ReRAM-based accelerators, PRIME~\cite{chi2016prime}, ISAAC~\cite{shafiee2016isaac}, and PipeLayer~\cite{song2017pipelayer}, especially PRIME (as detailed information is available).
Previous studies already show great speedup over conventional digital circuits.
For example, Eyeriss~\cite{chen2016eyeriss} achieves $35$ frame/s throughput and $115.4ms$ latency for AlexNet on a chip of $12.25mm^2$ under $65nm$ process \textit{with} off-chip memory, while we achieve $28.2$K frame/s and $100.49\mu s$ on $51.86mm^2$ under $45nm$ process \textit{without} off-chip memory.
Most of the improvements come from device benefit.
Thus, we only compare with ReRAM-based accelerators to show the improvements from the innovation at the architecture and system levels.

\textbf{FPSA Configuration.}
The crossbar size is set to $256\times512$; the positive and negative values of each logic column is represented with two adjacent crossbar-columns respectively.
Logically, the crossbar size is $256\times256$.
At each intersection, we put $8$ cells connected in parallel.
Each cell can be set to $16$ levels ($4$-bit), and we add up the values of $8$ cells to represent an $8$-bit weight.
This is done for reliability reasons, which will be discussed in Section~\ref{sec:variation}.
We integrate $128$ LUTs in one CLB to make the area and number of pins of one CLB similar to one PE.
For SMBs, we choose SRAM with $16$Kb capacity.

\textbf{Simulation Setup.}
We use mrVPR tool for mrFPGA~\cite{cong2011mrfpga} as the placement \& routing tool to evaluate the area consumption and critical path for communication.
The mrVPR has two inputs: one is an architecture description file that contains the parameters of all the function blocks, and the other is a netlist composed of these blocks.
We implement the neural synthesizer to generate the core-op graph and the spatial-to-temporal mapper to generate the function-block netlist for mrVPR.
The parameters of function blocks are listed in Table~\ref{tab:fpsa_parameters}.
We use NVSim~\cite{dong2014nvsim} to evaluate ReRAM-crossbar, sense amplifier, SMB and CLB, and use Synopsys Design Compiler for other peripheral circuits; all are under the $45nm$ process.
The routing architecture is stacked over function blocks. According to the report from mrVPR, the area of the former is less. 
We build a simulator to evaluate the performance based on the reported routing result from mrVPR.

\textbf{Methodology.}
To show the effects of the new routing architecture and simplified PEs, we first compare PRIME with FP-PRIME (FPSA's routing architecture with PRIME's PE) to show that the \textit{communication bound} of PRIME can be broken.
Then, FP-PRIME is compared with FPSA to show the further improvement from the new PE circuits.
In addition, we evaluate FPSA with different models to give the overall performance.

\subsection{Performance Improvement}


\textbf{Overall Comparison.}
In Figure~\ref{fig:overall_comparison}, we compare PRIME, FP-PRIME, and FPSA for VGG16.
FP-PRIME is composed of the FPSA routing architecture and PRIME's PEs, whose \textit{peak performance} and \textit{ideal performance} are the same as PRIME's.
The performance improvements comes from the three aspects listed in Section~\ref{sec:motivation}: \textit{Improving Communication}, \textit{Reducing Area}, \textit{Reducing Latency}.

\noindent$\bullet$ \textit{Improved Communication.}
Comparing PRIME and FP-PRIME in Figure~\ref{fig:overall_comparison}, we can see that by introducing the reconfigurable routing architecture, FP-PRIME can break the \textit{communication bound}.
Its performance is very close to the ideal case (the gap looks negilible in the logarithmic axes).

\noindent$\bullet$ \textit{Reduced Area \& Latency.}
Comparing FP-PRIME and FPSA, we can further increase the performance due to the area \& latency reduction of our PE design.

\noindent Combining these together, we can achieve up to $1000\times$ speedup with the same area consumption.

\textbf{Communication Improvement.}
In Figure~\ref{fig:latency_breakdown}, we show the average latency of computation and communication of one PE for VGG16.
The communication takes most of latency of PRIME.
By introducing the reconfigurable routing, the communication latency is reduced to $59.4 ns$, which is negilible compared to the computation time, $3064.7 ns$.
By further simplifying the peripheral circuits of PE, the computation time is reduced to $156.4 ns$, while the communication time increases to $633.9 ns$ because we transmit the spike trains directly instead of spike counts.
The communication overhead is simply the reason for the gap between the ideal case and the real case for FPSA in Figure~\ref{fig:overall_comparison}.
It can be improved by adding buffers:
Currently, the input spike signal of the charging unit is hold by its source PE.
If we add more buffers between the source and target PEs, the latency could be reduced, but it will also decrease the density advantage of current FPSA design.
We will discuss more about the effect of transmitting spike trains in Section~\ref{sec:spiking}.

\textbf{Area \& Latency Reduction.}
In Table~\ref{tab:ops_comprison}, we compare the area and latency of one PE in PRIME and those in FPSA.
The area is reduced by $36.63\%$ and the latency is reduced by $94.90\%$, which leads to the overall improvement on computational density by $31\times$.
The major improvements are from latency reduction because we do not need to share simplified peripheral circuits among different rows and columns.
The computational density is $38.004TOPS/mm^2$, which is higher than PRIME~\cite{chi2016prime} ($1.229TOPS/mm^2$), PipeLayer~\cite{song2017pipelayer} ($1.485TOPS/mm^2$), and ISAAC~\cite{shafiee2016isaac} ($0.479TOPS/mm^2$).

\subsection{Scalability \& Utilization}
We test the performance of FPSA under $1\times$, $4\times$, $16\times$, and $64\times$ \textit{duplication degrees} (defined in Section~\ref{sec:mapper}) for all the benchmark models, results in Figure~\ref{fig:scalability}.
The detailed performance for the $64\times$ case is listed in Table~\ref{tab:fpsa_overall}.

In Figure~\ref{fig:fpsa_performance}, with $4\times$, $16\times$, and $64\times$ \textit{duplication degree}, the geometric mean of the performance improvement is $3.06\times$, $10.88\times$, and $38.65\times$, respectively.
In contrast, the increase of the geometric mean of area consumption is only $1.25\times$, $1.85\times$, and $3.73\times$, respectively.
Especially, for the last four ImageNet models, the area consumption is only increased by $1.003\times$, $1.074\times$, and $1.504\times$ on average.

The reason for the super-linear scalability is the increased utilization when more resources are available.
In Figure~\ref{fig:fpsa_utilization}, we show the peak computational density, the spatial utilization bound (due to the imperfect crossbar mapping), the temporal utilization bound (due to the unbalanced workload), and the real computational density.
The two bounds depend on the property of the models:
There is no weight sharing in the MLP model, so its workload is balanced and the two bounds coincide with each other.
For CNN models, when more resources are available, the spatial utilization bounds do not change (we will discuss how to improve this bound in Section~\ref{sec:utilization}).
But the temporal utilization bound will increase significantly, which provides the super-linear scalability (as long as the communication bound is not hit)

\section{Discussion}\label{sec:discussion}
Despite overall improvements, there are also some other considerations that affect our design details.

\subsection{Spiking Schema}\label{sec:spiking}
Spiking schema has been used in existing design, e.g. PipeLayer~\cite{song2017pipelayer}, to reduce the overhead of ADC and DACs, but there is a significant different between our work and theirs.
We transmit spike trains directly through the routing architecture while they transmit the spike counts.
Despite the saved overhead of encoder/decoder circuits, it can also reduce end-to-end latency and on-chip buffers.

As discussed in Section~\ref{sec:mapper}, when two PEs are connected directly without buffers, the post-PE can start computation only $1$ cycle after the pre-PE starts (the No-Buffer Dependency (NBD)), and we only need 1-bit buffer to store current spike.
If we transmit the spike counts, the post-PE should wait for at least $2^n$ cycles (the sampling window for $n$-bit number) until the pre-PE finish all its computation, and then start its computation. 
In addition, it needs $n$-bit buffer to store the spike count.
Thus, by transmitting the spike trains directly, we can gain up $2^n\times$ end-to-end latency reduction for NBD and $n\times$ buffer consumption saving.
The drawback is that we will generate $2^n$-bit traffic for an $n$-bit number, which is the reason for the increased communication latency from FP-PRIME to FPSA in Figure~\ref{fig:latency_breakdown}.
But compared to the original latency of PRIME, it is negligible.
We list them in Table~\ref{tab:fpsa_overall}: the latency for VGG16 is only $671.8\mu s$ while PRIME's is $102.0ms$.

\subsection{Device Variation and NN accuracy}\label{sec:variation}
ReRAM devices are not ideal.
Due to the programming overhead and the intrinsic working mechanism of ReRAM cells, its conductance value cannot be programmed to the exact value as expected; the conductance value also has cycle-to-cycle variation~\cite{yao2017rram}.
The device variation will inevitably lead to inaccurate results even if we set a tight margin between levels.
The reason is that, in the ReRAM-crossbar based computing, there is no explicit read to quantize the obtained conductance, and all currents (with errors) from cell along the same column will accumulate.
Some software approaches, e.g. Vortex~\cite{liu2015vortex}, have been proposed to make NN models more robust to variation.
We have adopted these methods in our neural synthesizer, but as the inherent fault tolerance of NN is limited, for relative large variation, the effect is limited.
Thus, from the architecture perspective, we should also leverage more cells for one weight value to reduce the variation exposed to software level.
Without loss of generality, suppose the conductance of an ReRAM cell is a random variable obeys a normal distribution $N(\mu, \sigma^2)$ rather than a number.
We use \textit{normalized deviation}, which is the ratio between the standard deviation and the value range, to measure the variation exposed to software.

\textbf{The existing splicing method.}
Most existing architecture studies~\cite{chi2016prime,shafiee2016isaac} employ the \textit{splicing method}, which uses multiple cells for different bits of a number, to increase the representation precision of ReRAM. Suppose we use two $n$-bit cells to form a number of $2n$-bit cell, one for the high $n$ bits and one for the low $n$ bits. Their conductance values are $H$ and $L$, respectively: $H\sim N(h, \sigma^2)$ and $L\sim N(l, \sigma^2)$ where $h$ and $l$ are the expected values of the high $n$ bits and low $n$ bits, respectively. The number should be expressed as $2^n H+L\sim N(2^n h+l, (2^n\sigma)^2 + \sigma^2)$.
Its \textit{normalized deviation} is $\sqrt{2^{2n}+1}\sigma/(2^{2n}-1)$, which is almost equal to the ratio of one-cell case, $\sigma/(2^n-1)$.
Namely, it has little improvement on accuracy.

\textbf{The new add method.}
We propose the \textit{add method} that will add the conductance values evenly to increase precision and reduce variation.
Considering the general case that $n$ cells ($X_1,\ldots,X_n$ and $X_i\sim N(x_i,\sigma^2)$) are joined together by coefficient $a_1,\ldots,a_n$. Then the number is expressed as $\sum_i a_i X_i\sim N(\sum_i a_i x_i, \sum_i (a_i\sigma)^2)$.
The \textit{normalized deviation} is decreased by $\sum_i |a_i|/\sqrt{\sum_i a_i^2}$.
According to Cauchy inequality, the deviation decrease would reach its maximum value $\sqrt{n}$ when $|a_1|=\ldots=|a_n|$.

\begin{figure}
    \centering
    \includegraphics[width=0.48\textwidth]{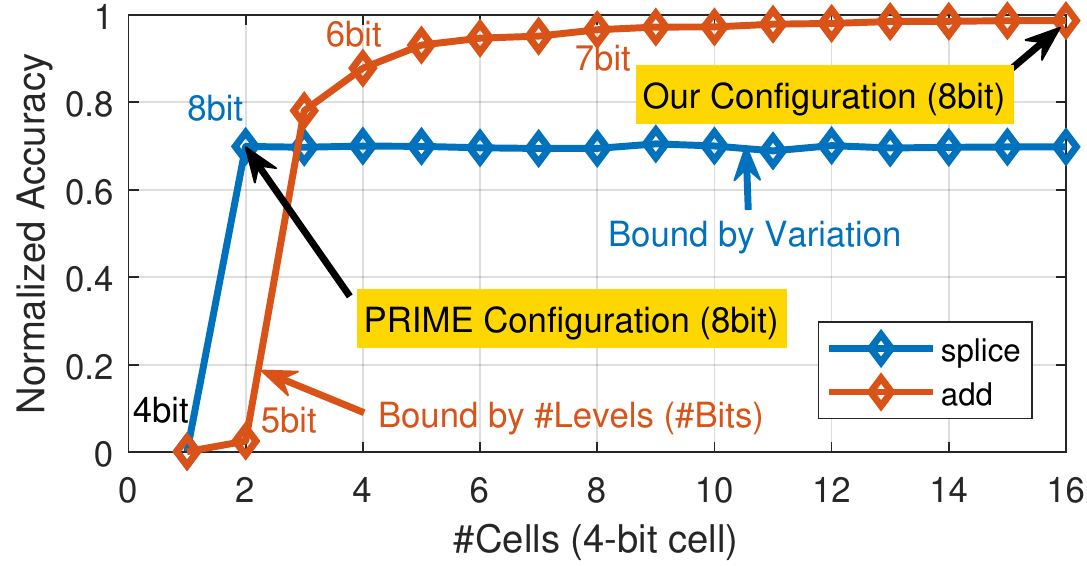}
    \caption{The normalized accuracy of VGG16 (normalized by the full precision accuracy) for the \textit{splice} and \textit{add} method with different number of cells used (4-bit for each cell).}
    \label{fig:splice_add}
    \vspace{-10pt}
\end{figure}

Figure~\ref{fig:splice_add} shows the effect of the two methods on the accuracy of VGG16.
The variation data is derived from real fabricated ReRAM cells~\cite{yao2017rram}.
PRIME use two 4-bit cells to form an 8-bit weight value with \textit{splicing}.
The accuracy drops to $70\%$ of the full precision accuracy.
In our design, we use 16 4-bit cells, 8 for positive and 8 for negative to form an 8-bit weight value with \textit{add}.
The accuracy is close to full precision accuracy.

\subsection{Spatial Utilization}\label{sec:utilization}
The \textit{Spatial Utilization Bound} comes from the fact that weight matrices cannot fit crossbars perfectly. Moreover, we find that the neural synthesizer aggravates this situation.
It introduces many small-scale weight matrices to implement operations such as reduction and max pooling.
For example, in GoogleNet, after synthesis the pooling operations occupy $67.2\%$ of PEs, which leads to the large gap between the \textit{peak performance} and the \textit{spatial utilization bound} in Figure~\ref{fig:fpsa_utilization}.
To improve the utilization, from the hardware perspective, we could introduce different scales of PE to fit weight matrices better.
From the software perspective, a future task is to find a better set of operations supported by hardware than the core-op.
\section{Conclusion}
By analyzing the bottlenecks and bounds for ReRAM-based NN acceleration, we propose a full system design of ReRAM-based NN accelerator, from the circuit level to the architectural and system level. 
Owing to the software system and massive hardware resources, it can support the function diversity and optimized execution of NN models on the proposed compact and efficient ReRAM PEs, achieving up to $1000\times$ speedup compared to an existing ReRAM-based design, PRIME. 
Last but not least, the computational density, $38TOPS/mm^2$, is also much higher than counterparts. 

%
\begin{acks}
Thanks for the support from Beijing Innovation Center for Future Chip, the support of the Science and Technology Innovation Special Zone project, China, and the support of HUAWEI project.
This work was also supported by NSF grant CCF 1500848, 1719160, 1725447, 1730309, 1740352, SRC nCORE NC2766-A, and CRISP, one of six centers in JUMP, a SRC program sponsored by DARPA.
\end{acks}

%
\bibliographystyle{ACM-Reference-Format}
\bibliography{references}

%

\end{document}